\documentclass[12pt]{article}
\usepackage{amssymb, amsmath, amsfonts}
\newtheorem{lemma}{Lemma}

\newtheorem{definition}{Definition}

\newtheorem{theorem}{Theorem}

\newtheorem{example}{Example}

\begin{document}

\begin{center}

{\Large Cartan matrices and integrable lattice Toda field equations}

\vskip 0.2cm

{Ismagil Habibullin}\footnote{e-mail: habibullinismagil@gmail.com}\\

{Ufa Institute of Mathematics, Russian Academy of Science,\\
Chernyshevsky Str., 112, Ufa, 450008, Russia}\\

\bigskip

{Kostyantyn Zheltukhin}\footnote{e-mail: zheltukh@metu.edu.tr}\\
{Department of Mathematics, Middle East Technical University, 06531 Ankara, Turkey}

\bigskip

{Marina Yangubaeva}\footnote{e-mail: marmarishka@list.ru}\\
{Faculty of Physics and Mathematics, 
 \\Birsk State Social Pedagogical Academy,\\
 Internationalnaya Str., 10, Birsk, 452452, Russia \\}

\end{center}

\begin{abstract}
Differential-difference integrable exponential type systems are studied corresponding to the Cartan matrices of semi-simple or affine Lie algebras. For the systems corresponding to the algebras $A_2$, $B_2$, $C_2$, $G_2$ the complete sets of integrals in both directions are found. For the simple Lie algebras of the classical series $A_N$, $B_N$, $C_N$ and affine algebras of series $D^{(2)}_N$  the corresponding systems are supplied with the Lax representation.
\end{abstract}



\section{Introduction}

It is well known (see \cite{MOP}) that one can assign an integrable exponential type system of hyperbolic equations to each semi-simple or affine Lie algebra. Denote through $A=\{a_{ij}\}$ the Cartan matrix, corresponding to the Lie algebra, then the desired exponential system is written as follows
\begin{equation}\label{cont_system1}
r_{x,y}^i=e^{\sum{a_{ij}r^j}}.
\end{equation}
The classical Toda field theories in the continuous space-time are based on infinite dimensional integrable Hamiltonian systems closely connected with (\ref{cont_system1}). In this context these equations have been intensively studied in mathematical physics during the last three decades \cite{MOP}-\cite{ShabatYamilov}.

In the present article we study the problem of discretization of integrable systems of the form (\ref{cont_system1}) preserving  integrability, it is important since discretization makes a bridge between quantum and classical field theories.  More exactly, we look for integrable differential-difference systems, with a discrete space-like variable $n$ and a continuous time $x$ which correspond to (\ref{cont_system1}) in the continuum limit. The problem of discretization of exactly solvable models is intensively studied. For the particular case $x=y$ Suris \cite{Suris} exhibited discrete integrable versions of the Toda chain corresponding
to all the non-exceptional affine Lie algebras. Today different approaches to the problem are known in literature \cite{Hirota}-\cite{Sakieva}. The approach using the Hamiltonian formulation of the Toda field equations is studied in \cite{IH}, where the equation of motion for the Toda lattice, associated with the simple Lie algebras, is derived. In order to approve integrability of the lattices suggested the authors studied the properties of the corresponding $\tau$-functions. Further discretization of the lattice model leads to difference-difference equations on discrete space-time, having a very large variety of applications in Physics, related
"... to commuting transfer matrices in solvable lattice models, q-characters of
Kirillov-Reshetikhin modules of quantum affine algebras, cluster algebras
with coefficients, periodicity conjectures of Zamolodchikov and others,
dilogarithm identities in conformal field theory, difference analog of L-operators
in KP hierarchy, Stokes phenomena in 1D Schr{\" o}dinger problem,
AdS/CFT correspondence, Toda field equations on discrete space-time, Laplace
sequence in discrete geometry, Fermionic character formulas and combinatorial
completeness of Bethe ansatz, Q-system and ideal gas with exclusion statistics,
analytic and thermodynamic Bethe ansatz, quantum transfer matrix method
and so forth\footnote{A. Kuniba, T.Nakanishi, J.Suzuki, {\it $T$-systems and $Y$-systems in integrable systems}, J. Phys. A: Math. Theor. 44 (2011) 103001 (146pp)}".   

Our approach to the problem of constructing discrete analogues of system (\ref{cont_system1}) is based on the technique of integrable cutting off constraints \cite{H}, \cite{KhG} and on the integrals preserving method of discretization, suggested recently in \cite{Sakieva},  for the Darboux integrable hyperbolic type models. 

Define a mapping assigning to each $N\times N$ matrix $A$ a system of differential-difference equations of the form
\begin{equation}\label{cont_system3}
r_{1,x}^i-r_{x}^i=e^{\sum{a_{i,j}^{+}r_1^j}+\sum{a_{i,j}^{-}r^j}},\quad i,j=1,2,...,N,
\end{equation}
where the given matrix $A$ is decomposed into a sum of two triangular matrices: $A=A_{+} + A_{-}$ with $A_{+}=\{a_{ij}^{+}\}$ being upper triangular and $A_{-}=\{a_{ij}^{-}\}$ - lower triangular matrices such that all diagonal entries of $A_{+}$ equal unity. Here the functions $r^{j} = r^{j}(n,x)$, $ j=1, ..., N$ are the searched field variables. Everywhere below the subindex denotes a shift of the discrete variable $n$ or the derivative with respect to $x$: $r_k^{j}=r^j(n+k,x)$ and $r_{[i]}^{j}=\frac{\partial^i}{\partial x^i}r^j(n,x)$. Sometimes we use also $r_x^{j}=\frac{\partial}{\partial x}r^j(n,x)$, $r_{xx}^{j}=\frac{\partial^2}{\partial x^2}r^j(n,x), ...\,$. We refer to (\ref{cont_system3}) as a differential-difference exponential type system.

The further investigations approve the following hypothesis.

\noindent
{\bf Conjecture 1.} 
a) If $A$ is the Cartan matrix of a semi-simple Lie algebra then (\ref{cont_system3}) is  Darboux integrable. More detailed explanation of Darboux integrability is given below in Definition \ref{def1} Roughly speaking Darboux integrable equations can be solved in quadratures.\\
b) If $A$ is the Cartan matrix of an affine Lie algebra then (\ref{cont_system3}) is $S$-integrable (it admits large classes of exact multi-soliton solutions). 

Part a) of the conjecture is checked for the following cases: $A_2$, $B_2$, $C_2$, $G_2$. In these cases the exponential system (\ref{cont_system3}) reads as follows
\begin{eqnarray} 
r^1_{1,x}-r^1_x&=&e^{r^1+r^1_1-r^2_1},\label{sysu} \\
r^2_{1,x}-r^2_x&=&e^{-cr^1+r^2+r^2_1},\nonumber
\end{eqnarray}
where $c=1, 2, 3$. Integrals in both directions for these systems are found.
For the systems related to the algebras of the series $A_N$, $C_N$ with $N\geq2$ the Lax pairs are given. They are also Darboux integrable.
Note that the systems corresponding to the Cartan matrices $C_N$ and $B_N$,   $N=2,3\dots$, are the same up to renaming the variables. The system corresponding to $A_N$ is easily reduced to that found earlier in \cite{Hirota}.

Part b) is illustrated by an example of system (\ref{cont_system3}) corresponding to the Cartan matrix of algebra $D^{(2)}_N$ for which the Lax pair is found by using the integrable cutting off constraint for infinite chain of the B\"acklund transformations for the classical two-dimensional Toda lattice.

Let us recall the necessary definitions (see \cite{ZhiberGur}). Function $w=w(r,r_{[1]},r_{[2]},...,r_{[k]})$ with $r=(r^1,r^2,...r^N)$, $r_{[i]}=\frac{\partial^i r}{\partial x^i}$ is called a $y$-integral of order $k$ of system (\ref{cont_system1}) if its total derivative with respect to $y$ vanishes $D_yw=0$. Similarly, function $\bar w=\bar w(r,\bar r_{[1]},\bar r_{[2]},...,\bar r_{[k]})$, where $\bar r_{[i]}=\frac{\partial^i r}{\partial y^i}$
is an $x$-integral of order $k$ if $D_x\bar w=0$. System (\ref{cont_system1}) is called Darboux integrable if it admits $N$ $y$-integrals $w^{(1)},w^{(2)},\ldots, w^{(N)}$ of orders $q^{(1)},q^{(2)},\ldots, q^{(N)}$ correspondingly and $N$ $x$-integrals $\bar w^{(1)},\bar w^{(2)},\ldots, \bar w^{(N)}$ of orders  $p^{(1)},p^{(2)},\ldots, p^{(N)}$ such that 
\begin{equation}\label{det}
det \left( \frac{\partial w^{(k)}}{\partial r^j_{[q^{(k)}]}}\right)\neq 0 \qquad \mbox{and} \qquad det \left( \frac{\partial \bar w^{(k)}}{\partial \bar r^j_{[p^{(k)}]}}\right)\neq 0, \quad k,j=1,2,...N.
\end{equation}
In this case the set of $2N$ above integrals constitute a basis of integrals of the system (\ref{cont_system1}). Integrals $w^{(1)},w^{(2)},\ldots, w^{(N)}$ (or integrals $\bar w^{(1)},\bar w^{(2)},\ldots, \bar w^{(N)}$) satisfying (\ref{det}) are called independent.

Analogously we define integrals for the differential-difference chain (\ref{cont_system3}). Here the set of dynamical variables consists of the field variables $r^1,r^2,\ldots, r^N$, their shifts $D^kr^j=r^j_k$ and derivatives $D^i_xr^j=r^j_{[i]}$. Recall that  $D$ is the shift operator acting as follows $Dh(n)=h(n+1)$ and $D_x$ is the operator of the total derivative with respect to $x$. For the convenience we put $r=(r^1,r^2,\ldots, r^N)$. Function $F=F(x,r_{k'},r_{k'+1},\ldots, r_k)$, where $k'\leq k$ is called an $x$-integral of system (\ref{cont_system3}) of order $k-k'$ if $D_x F=0$. And $I=I(x,r,r_{[1]},r_{[2]},\ldots ,r_{[k]})$ is called an $n$-integral of order $k$ if $DI=I$.

\begin{definition}\label{def1}
 We call system (\ref{cont_system3}) Darboux integrable if it admits $N$ $x$-integrals $F^{(1)},F^{(2)},\ldots, F^{(N)}$ of the form $F^{(i)}=F^{(i)}(x,r,r_1,\ldots,r_k)$ of order $k$ and $N$ $n$-integrals $I^{(1)},I^{(2)},\ldots,I^{(N)}$ of order $s$ such that 
\begin{equation}\label{det2}
det \left( \frac{\partial F^{(i)}}{\partial r^j_{k}}\right)\neq 0 \qquad \mbox{and} \qquad det \left( \frac{\partial I^{(i)}}{\partial r^j_{[s]}}\right)\neq 0,\quad i,j=1,2,\dots N.
\end{equation}
\end{definition}
A set of $2N$ integrals satisfying (\ref{det2}) constitutes a basis of integrals of system (\ref{cont_system3}). Integrals $F^{(1)},F^{(2)},\ldots, F^{(N)}$ as well as integrals $I^{(1)},I^{(2)},\ldots,I^{(N)}$ satisfying (\ref{det2}) are called independent because non of them can be functionally expressed through the other integrals and their shifts and derivatives.

\noindent

{\bf Conjecture 2.} Suppose that the function 
$$w=w(r,r_x,r_{xx},...)$$
is a $y$-integral for the system (\ref{cont_system1}) with the Cartan matrix $A$ of a semi-simple Lie algebra. Then this function is also an $n$-integral for the system (\ref{cont_system3}) with the same matrix $A$.

Conjecture 2 is approved by the examples below (see section 3). For instance, the functions 
\begin{eqnarray}
I^{(1)}&=& w^{(1)} = r^1_{xx}+r^2_{xx}-(r^1_x)^2+r^1_xr^2_x-(r^2_x)^2,\nonumber\\
I^{(2)}&=& w^{(2)}=r^1_{xxx}+r^1_x(r^2_{xx}-2r^1_{xx})+(r^1_x)^2r^2_x-r^1_x(r^2_x)^2\nonumber
\end{eqnarray}
are $y-$integrals for the system
\begin{eqnarray}
r^1_{xy}=e^{2r^1-r^2},\nonumber\\
r^2_{xy}=e^{-r^1+2r^2}\nonumber
\end{eqnarray}
and $n-$integrals for the differential-difference chain
\begin{eqnarray}
r^1_{1,x}-r^1_x&=&e^{r^1+r^1_1-r^2_1},\nonumber\\ 
r^2_{1,x}-r^2_x&=&e^{-r^1+r^2+r^2_1}.\label{A-chain}
\end{eqnarray}
A set of independent $x$-integrals for chain (\ref{A-chain}) is given in Example \ref{x-int_c=1}  

To show the existence of $x$- and $n$-integrals one can consider characteristic $x$-algebra $L_x$ and characteristic $n$-algebra $L_n$  (see \cite{HP} and references therein). 

\begin{theorem} (see \cite{HP} ) 
A system of equations admits a non trivial $x$-integral if and only if its characteristic $x$-algebra is of finite dimension.
\end{theorem}

\begin{theorem} (see \cite{HP}) 
A system of equations admits a non trivial $n$-integral if and only if its characteristic $n$-algebra is of finite dimension.
\end{theorem}

\noindent
{\bf Conjecture 3.} Suppose that the evolutionary type system
$$
r_\tau=f(r,r_x,r_{xx},\dots)
$$
is a symmetry of system (\ref{cont_system1}) with the Cartan matrix of a semi-simple or affine Lie algebra. Then this equation is also a symmetry of system (\ref{cont_system3}) with the same matrix $A$.

The paper is organized as follows. In section 2 we introduce characteristic $x$-algebra for a system of the form (\ref{sys}) (see also (\ref{sysu})) and prove that for the values of $c=1,2,3$ the algebra is of finite dimension. This allows one to deduce a system of linear first order PDE the $x$-integrals should satisfy to. These systems of equations are successfully solved and independent integrals are found for each of the cases $c=1,2,3$. In the third section $n$-integrals for system (\ref{sys}) are presented. The close connection of the integrals found with $y$-integrals for the corresponding continuous systems (\ref{cont_system1}) is demonstrated. 

In fourth section an alternative approach is considered to construct 
finite field systems. In this approach one uses the cutting off 
boundary conditions preserving integrability for the open chains (see 
\cite{H}, \cite{KhG}).
By using certain special choices of cutting off constraint we have been able to reproduce in this way some of the chains obtained by applying the mapping above (see (\ref{cont_system3})).
The second method to construct integrable systems is also important since it allows one to find the Lax pairs for the finite field reductions.

\section{Characteristic $x$-algebras and $x$-integrals}

Let us  consider the problem of finding the  $x$-integrals for the system (\ref{sysu}) which is rewritten for the convenience in the following form
\begin{eqnarray} 
a_{1,x}-a_x&=&e^{a+a_1-b_1},\label{sys} \\
b_{1,x}-b_x&=&e^{-ca+b+b_1},\nonumber
\end{eqnarray}
where $c=1,2,3$, $a=r^1$, $b=r^2$. First we introduce the notion of the characteristic $x$-algebra for the chain (\ref{sys}). According to the definition an $x$-integral $F$ should satisfy the equation $D_xF(x,a,b,a_1,b_1,a_{-1},$ $b_{-1},...)=0$ which is equivalent to the equation $ZF=0$ where 
\begin{equation}\label{Z}
Z=\frac{\partial}{\partial x}+\sum_{k=-\infty}^{\infty} \left(a_{k,x}\frac{\partial}{\partial a_k}+
b_{k,x}\frac{\partial}{\partial b_k}\right).
\end{equation}
Another important observation is that $x$-integral cannot depend on the variables $a_x$, $b_x$ despite the coefficients of the equation $ZF=0$ depend on these variables. That is why we have to put two additional equations $Y_1F=0$, $Y_2F=0$ where 
$$Y_1=\frac{\partial}{\partial a_x}\,\quad
\mbox{and} \quad\, Y_2=\frac{\partial}{\partial b_x}.
$$
Denote through $L_x$ the Lie algebra generated by the operators $Z$, $Y_1$, $Y_2$ over the field of constants. We call this algebra characteristic Lie algebra in $x$-direction or $x$-algebra (see also \cite{HP}). The purpose of the present section is to give the complete description of the $x$-algebra for system (\ref{sys}).

For the system (\ref{sys}) we have 
\begin{eqnarray}
a_{k,x}=a_x+\sum_{i=1}^{k} e^{a_{i-1}+a_i-b_i} \quad \mbox{and} \quad
a_{-k,x}=a_x-\sum_{i=1}^{k} e^{a_{-i}+a_{-i+1}-b_{-i+1}},\nonumber\\
b_{k,x}=b_x+\sum_{i=1}^{k} e^{-ca_{i-1}+b_{i-1}+b_i} \quad \mbox{and} \quad
b_{-k,x}=b_x-\sum_{i=1}^{k} e^{-ca_{-i}+b_{-i}+b_{-i+1}}.\nonumber
\end{eqnarray}
So, the vector field $Z$ takes a form
\begin{multline}
Z=\frac{\partial}{\partial x}+a_x\frac{\partial}{\partial a}+b_x\frac{\partial}{\partial b}+\\
+\sum_{k=1}^{\infty} \left(a_x+\sum_{i=1}^{k} e^{a_{i-1}+a_i-b_i}\right)\frac{\partial}{\partial a_k}+
\sum_{k=1}^{\infty} \left(a_x-\sum_{i=1}^{k} e^{a_{-i}+a_{-i+1}-b_{-i+1}}\right)\frac{\partial}{\partial a_{-k}} + \\
+\sum_{k=1}^{\infty} \left(b_x+\sum_{i=1}^{k} e^{-ca_{i-1}+b_{i-1}+b_i}\right)\frac{\partial}{\partial b_k}+
\sum_{k=1}^{\infty} \left(b_x-\sum_{i=1}^{k} e^{-ca_{-i}+b_{-i}+b_{-i+1}}\right)\frac{\partial}{\partial b_{-k}}.
\end{multline}
Taking commutators of the vector fields $Y_1$, $Y_2$ and $Z$ we get
$$A=[Y_1, Z]=\sum_{k=-\infty}^{\infty} \frac{\partial}{\partial a_k}, \quad
B=[Y_2, Z]=\sum_{k=-\infty}^{\infty} \frac{\partial}{\partial b_k}.$$
Taking commutators of the vector fields $A$, $B$ and $Z$ we get vector fields $A^*=[A, Z]$ and $B^*=[B, Z]$:
\begin{multline}
A^*=\sum_{k=1}^{\infty} \left(2\sum_{i=1}^{k} e^{a_{i-1}+a_i-b_i}\right)\frac{\partial}{\partial a_k}+
\sum_{k=1}^{\infty} \left(-2\sum_{i=1}^{k} e^{a_{-i}+a_{-i+1}-b_{-i+1}}\right)\frac{\partial}{\partial a_{-k}} + \\
+\sum_{k=1}^{\infty} \left(-c\sum_{i=1}^{k} e^{-ca_{i-1}+b_{i-1}+b_i}\right)\frac{\partial}{\partial b_k}+
\sum_{k=1}^{\infty} \left(c\sum_{i=1}^{k} e^{-ca_{-i}+b_{-i}+b_{-i+1}}\right)\frac{\partial}{\partial b_{-k}}
\end{multline}
and
\begin{multline}
B^*=\sum_{k=1}^{\infty} \left(-\sum_{i=1}^{k} e^{a_{i-1}+a_i-b_i}\right)\frac{\partial}{\partial a_k}+
\sum_{k=1}^{\infty} \left(\sum_{i=1}^{k} e^{a_{-i}+a_{-i+1}-b_{-i+1}}\right)\frac{\partial}{\partial a_{-k}} + \\
+\sum_{k=1}^{\infty} \left(2\sum_{i=1}^{k} e^{-ca_{i-1}+b_{i-1}+b_i}\right)\frac{\partial}{\partial b_k}+
\sum_{k=1}^{\infty} \left(-2\sum_{i=1}^{k} e^{-ca_{-i}+b_{-i}+b_{-i+1}}\right)\frac{\partial}{\partial b_{-k}}.
\end{multline}
It is convenient to introduce the following vector fields
\begin{multline}
P_1=\frac {1}{4-c} \left(2A^*+cB^*\right)=\\
=\sum_{k=1}^{\infty} \left(\sum_{i=1}^{k} e^{a_{i-1}+a_i-b_i}\right)\frac{\partial}{\partial a_k}+
\sum_{k=1}^{\infty} \left(\sum_{i=1}^{k} e^{a_{-i}+a_{-i+1}-b_{-i+1}}\right)\frac{\partial}{\partial a_{-k}} 
\end{multline}
and
\begin{multline}
P_2=\frac {1}{4-c} \left(A^*+2B^*\right)=\\
=\sum_{k=1}^{\infty} \left(\sum_{i=1}^{k} e^{-ca_{i-1}+b_{i-1}+b_i}\right)\frac{\partial}{\partial b_k}+
\sum_{k=1}^{\infty} \left(-2\sum_{i=1}^{k} e^{-ca_{-i}+b_{-i}+b_{-i+1}}\right)\frac{\partial}{\partial b_{-k}}. 
\end{multline}

Obviously the vector fields                                 
$$ \frac{\partial}{\partial x},\quad Y_1, \quad Y_2,\quad A,\quad B,\quad P_1 \quad \mbox{and} \quad  P_2$$\\
belong to the $x$-algebra and vector field $Z$ can be obtained as their linear combination.  
We use the following lemma to show that $x$-algebra is of finite dimension.
\begin{lemma} \label{K=0_lemma}(see \cite{HP})
Suppose that the vector field
\begin{equation}
K=\sum_{k=1}^\infty \left( \alpha_k \frac{\partial}{\partial a_k} + \alpha_{-k} \frac{\partial}{\partial a_{-k}} \right) + \sum_{k=1}^\infty \left( \beta_k \frac{\partial}{\partial b_k} + \beta_{-k} \frac{\partial}{\partial b_{-k}} \right)
\end{equation}
satisfies the equality $DKD^{-1}=hK$, where $h$ is a function depending on shifts and derivatives of variables $a$ and  $b$, then $K=0$.  
\end{lemma}

Let us consider the sequence of commutators $T_0=P_2,\quad T_1=[P_1,P_2]$ and $T_n=[P_1,T_{n-1}]$, $n=1,2\dots$. To apply the lemma 1 we need 
to know how the transformation $D(*)D^{-1}$ acts on the vector fields  $A$, $B$, $P_1$ and $P_2$. One can easily check that
\begin{equation}
DAD^{-1}=A,\quad DBD^{-1}=B.
\end{equation}
 
\begin{lemma}
\begin{equation}
\begin{array}{l}
DP_1D^{-1}=P_1-e^{a+a_1-b_1}A,\\
DP_2D^{-1}=P_2-e^{-ca+b+b_1}B.\\
\end{array}
\end{equation} 
\end{lemma} 
{\bf Proof.} 
Let 
$\displaystyle{
DP_1D^{-1}=\sum_{k=1}^{\infty} \left(\mu_{k}\frac{\partial}{\partial a_{k}} +\mu_{-k}\frac{\partial}{\partial a_{-k}}\right), 
}$
then for $k=1,2\dots$
$$
\mu_k=DP_1D^{-1}(a_k)=DP_1(a_{k-1})=D\left(\sum_{i=1}^{k-1} e^{a_{i-1}+a_i-b_i}\right)=\sum_{i=1}^{k-1} e^{a_{i}+a_{i+1}-b_{i+1}}.
$$
Hence we have 
$$
\mu_k=\left(\sum_{i=1}^{k} e^{a_{i-1}+a_i-b_i}\right)-e^{a+a_1-b_1}.
$$
Similar equality holds for $\mu_{-k}$, $k=1,2\dots$. Thus, we have $DP_1D^{-1}=P_1-e^{a+a_1-b_1}A$.
Formula for $DP_2D^{-1}$ is proved in the same way. $\Box$

\begin{lemma}
For any $k\geq c+1$, where $c$ takes the values $c=1,2,3$ we have $T_k=0$.
\end{lemma}
{\bf Proof.}
Using previous lemma we have
\begin{multline}
DT_1D^{-1}= D[P_1,P_2]D^{-1}=[DP_1D^{-1},DP_2D^{-1}]=[P_1-e^{a+a_1-b_1}A,P_2-e^{-ca+b+b_1}B]= \\
T_1+ce^{a+a_1-b_1}P_2-e^{-ca+b+b_1}P_1-  ce^{(1-c)a+a_1+b}B.
\end{multline}
In the same way one can show 
\begin{multline}
DT_2D^{-1}=T_2+2(c-1)e^{a+a_1-b_1}T_1+c(c-1)e^{2a+2a_1-2b_1}P_2-\\
-2(c-1)e^{(1-c)a+a_1+b}P_1-c(c-1)e^{(2-c)a+2a_1+b-b_1}B,
\end{multline}
\begin{multline}
DT_3D^{-1}=T_3+3(c-2)e^{a+a_1-b_1}T_2+3(c-1)(c-2)e^{2a+2a_1-2b_1}T_1+\\
+c(c-1)(c-2)e^{3a+3a_1-3b_1}P_2-3(c-1)(c-2)e^{(2-c)a+2a_1+b-b_1}P_1-\\
-c(c-1)(c-2)e^{(3-c)a+3a_1+b-2b_1}B,
\end{multline}
\begin{multline}
DT_4D^{-1}=T_4+4(c-3)e^{a+a_1-b_1}T_3+6(c-2)(c-3)e^{2a+2a_1-2b_1}T_2+\\
+4(c-1)(c-2)(c-3)e^{3a+3a_1-3b_1}T_1+c(c-1)(c-2)(c-3)e^{4a+4a_1-4b_1}P_2-\\
-4(c-1)(c-2)(c-3)e^{(3-c)a+3a_1+b-2b_1}P_1-c(c-1)(c-2)(c-3)e^{(4-c)a+4a_1+b-3b_1}B.
\end{multline}
Hence the statement of the lemma directly follows from the lemma \ref{K=0_lemma} $\Box$

\begin{lemma}
For  $c=1,2,3$ we have $[P_2,T_1]=0$.
\end{lemma} 
{\bf Proof.} Direct calculations show that
$$
D[P_2,T_1]D^{-1}=[P_2,T_1] ,
$$
which implies that $[P_2,T_1]=0$ by lemma \ref{K=0_lemma} $\Box$

The following theorem can be easily proved by using the lemmas above.
\begin{theorem}
The $x$-algebra corresponding to system (\ref{sys}) with $c=1,2,3$ is of finite dimension.
\end{theorem}
{\bf Remark 1.} Emphasize that we do not study any classification problem for system (\ref{sys}) with arbitrary value of the parameter $c$, only three particular cases $c=1,2,3$. The problem of describing all values of $c$ for which $x$-algebra is of finite dimension is a subject of a separate investigation.

Since the $x$-algebra is finite-dimensional, we can find the $x$-integrals corresponding to the values of $c=1,2,3$ in system (\ref{sys}).

\begin{example}\label{x-int_c=1} 
For $c=1$ the corresponding $x$-algebra is generated by vector fields $\frac{\partial}{\partial x}$, $Y_1$, $Y_2$, $A$, $B$, $P_1$, $P_2$, $T_1$ with the following commutators table: 
\begin{center}
\begin{tabular}{|r|r|r|r|r|r|r|r|r|}
\hline
&$\frac{\partial}{\partial x}$&$Y_1$&$Y_2$&$A$&$B$&$P_1$&$P_2$&$T_1$\\
\hline
$\frac{\partial}{\partial x}$& $0$&$0$&$0$&$0$&$0$&$0$&$0$&$0$\\
\hline
$Y_1$& $0$&$0$&$0$&$0$&$0$&$0$&$0$&$0$\\
\hline
$Y_2$& $0$&$0$&$0$&$0$&$0$&$0$&$0$&$0$\\
\hline
$A$&$0$&$0$&$0$&$0$&$0$&$2P_1$&$-P_2$&$T_1$\\
\hline
$B$&$0$&$0$&$0$&$0$&$0$&$-P_1$&$2P_2$&$T_1$\\
\hline
$P_1$&$0$&$0$&$0$&$-2P_1$&$P_1$&$0$&$T_1$&$0$\\
\hline
$P_2$&$0$&$0$&$0$&$P_2$&$-2P_2$&$-T_1$&$0$&$0$\\
\hline
$T_1$&$0$&$0$&$0$&$-T_1$&$-T_1$&$0$&$0$&$0$\\
\hline
\end{tabular}
\end{center}

The $x$-integral 
$F(...a,b,a_1,b_1,a_2,b_2\dots)$ must satisfy equations
\begin{equation}
 A(F)=0,\quad B(F)=0,\quad P_1(F)=0,\quad P_2(F)=0,\quad T_1(F)=0. 
\end{equation}

Solving the above system, it is enough to assume that $F_1$ depends on $b,\,a,\,b_1,\,a_1,\,b_2,\,a_2$ and $F_2$ depends on $a,\,b_1,\,a_1,\,b_2,\,a_2,\,b_3$, we find two independent $x$-integrals
$$
\begin{array}{l}
F_1=e^{-b+b_1} + e^{-a+a_1+b_1-b_2} + e^{a_1-a_2},\\
F_2=e^{-a+a_1} + e^{a_1-a_2-b_1+b_2} + e^{b_2-b_3}.\\

\end{array}
$$

The system corresponds to the Cartan matrix 
\begin{equation}\label{Weil1}
N=\left( 
\begin{array}{cc}
2&-1\\
-1&2
\end{array}
\right)
\end{equation}
and the Cartan matrix  (\ref{Weil1}) is canonically related  with the simple Lie algebra $A_2$.\\
Recall that simple Lie algebra of rank $r$ is completely described by its Weyl generators $x_i, y_i,h_i,\quad 1\leq i \leq r$, such that 
\begin{enumerate}
	\item $x_i\neq 0$, $y_i\neq 0$, $h_i\neq 0$ for any $i$;
	\item for any $i,j$ the relations hold
	$$[h_i,h_j]=0,$$
	$$[x_i,y_j]=\delta_{ij}h_i,$$
	$$[h_i,x_j]=N_{ij}x_j,$$
	$$[h_i,y_j]=-N_{ij}y_j.$$
\end{enumerate}
Here $N=\{N_{ij}\}$ is the Cartan matrix of the algebra, and $\delta_{ij}$ is the Kroneker symbol.

It is evident from the table of commutators that the mapping $h_1\rightarrow-A$, $h_2\rightarrow-B$, $y_1\rightarrow P_1$, $y_2\rightarrow P_2$ gives an isomorphism between the algebra generated by the operators $\{A,B,P_1,P_2\}$ and Borel subalgebra of the Lie algebra $A_2$ generated by $\{h_1,h_2,y_1,y_2\}$.
\end{example}
       
\begin{example} 
For $c=2$ the corresponding $x$-algebra is generated by vector fields $\frac{\partial}{\partial x}$, $Y_1$, $Y_2$, $A$, $B$, $P_1$, $P_2$, $T_1$, and $T_2$ with the following commutators table:
\begin{center}
\begin {tabular}{|r|r|r|r|r|r|r|r|r|r|}
\hline
&$\frac{\partial}{\partial x}$&$Y_1$&$Y_2$&$A$&$B$&$P_1$&$P_2$&$T_1$&$T_2$\\
\hline
$\frac{\partial}{\partial x}$& $0$&$0$&$0$&$0$&$0$&$0$&$0$&$0$&$0$\\
\hline
$Y_1$& $0$&$0$&$0$&$0$&$0$&$0$&$0$&$0$&$0$\\
\hline
$Y_2$& $0$&$0$&$0$&$0$&$0$&$0$&$0$&$0$&$0$\\
\hline
$A$&$0$&$0$&$0$&$0$&$0$&$2P_1$&$-2P_2$&$0$&$2T_2$\\
\hline
$B$&$0$&$0$&$0$&$0$&$0$&$-P_1$&$2P_2$&$T_1$&$0$\\
\hline
$P_1$&$0$&$0$&$0$&$-2P_1$&$P_1$&$0$&$T_1$&$T_2$&$0$\\
\hline
$P_2$&$0$&$0$&$0$&$2P_2$&$-2P_2$&$-T_1$&$0$&$0$&$0$\\
\hline
$T_1$&$0$&$0$&$0$&$0$&$-T_1$&$-T_2$&$0$&$0$&$0$\\
\hline
$T_2$&$0$&$0$&$0$&$-2T_2$&$0$&$0$&$0$&$0$&$0$\\
\hline
\end{tabular}
\end{center}

The $x$-integral $F(...a,b,a_1,b_1,a_2,b_2\dots)$ must satisfy equations
\begin{equation}
\begin{array}{lll}
  A(F)=0, & B(F)=0, & P_1(F)=0, \\
\label{system2}\\
 P_2(F)=0, & T_1(F)=0,  & T_2(F)=0. \\
\end{array}  
\end{equation}

Solving system (\ref{system2}), it is enough to assume that $F_1$ depends on $a,\,b_1,\,a_1,\,b_2,\, a_2,\, b_3, \, a_3$ and $F_2$ depends on $b,\,a,\,b_1,\, a_1,\, b_2, \, a_2,\,b_3$, we find two independent $x$-integrals
$$
\begin{array}{l}
F_1=e^{-a+a_1} + e^{-a_1+a_2+b_2-b_3} + e^{a_1-a_2-b_1+b_2} + e^{a_2-a_3},\\
F_2= e^{-b+b_1} + e^{-2a+2a_1+b_1-b_2} + 2e^{-a+2a_1-a_2} + e^{2a_1-2a_2-b_1+b_2} + e^{b_2-b_3}.
\end{array}
$$ 

In this case 
$
N=\left( 
\begin{array}{cc}
2&-1\\
-2&2
\end{array}
\right)
$
and we use the map $h_1\rightarrow -B$, $h_2\rightarrow-A$,  $y_1\rightarrow P_2$, $y_2\rightarrow P_1$ to identify subalgebra generated by $\{A,B,P_1,P_2\}$ and the Borel subalgebra of $B_2$ generated by $\{h_1,h_2,y_1,y_2\}$. 
\end{example}

\begin{example} 
For $c=3$ the corresponding $x$-algebra is generated by vector fields $\frac{\partial}{\partial x}$, $Y_1$, $Y_2$, $A$, $B$, $P_1$, $P_2$, $T_1$, $T_2$, $T_3$ and $Q=[T_1,T_2]$ with the following commutators table:
\begin{center}
\begin{tabular}{|r|r|r|r|r|r|r|r|r|r|r|r|}
\hline
  &$\frac{\partial}{\partial x}$&$Y_1$&$Y_2$&$A$&$B$&$P_1$&$P_2$&$T_1$&$T_2$&$T_3$&$Q$\\
\hline
$\frac{\partial}{\partial x}$& 0&0&0&0&0&0&0&0&0&0&$0$\\
\hline
$Y_1$& $0$&$0$&$0$&$0$&$0$&$0$&$0$&$0$&$0$&$0$&$0$\\
\hline
$Y_2$& $0$&$0$&$0$&$0$&$0$&$0$&$0$&$0$&$0$&$0$&$0$\\
\hline
$A$&$0$&$0$&$0$&$0$&$0$&$2P_1$&$-3P_2$&$-T_1$&$T_2$&$3T_3$&$0$\\
\hline
$B$&$0$&$0$&$0$&$0$&$0$&$-P_1$&$2P_2$&$T_1$&$0$&$-T_3$&$Q$\\
\hline
$P_1$&$0$&$0$&$0$&$-2P_1$&$P_1$&$0$&$T_1$&$T_2$&$T_3$&$0$&$0$\\
\hline
$P_2$&$0$&$0$&$0$&$3P_2$&$-2P_2$&$-T_1$&$0$&$0$&$0$&$0$&$0$\\
\hline
$T_1$&$0$&$0$&$0$&$T_1$&$-T_1$&$-T_2$&$0$&$0$&$Q$&$0$&$0$\\
\hline
$T_2$&$0$&$0$&$0$&$-T_2$&$0$&$-T_3$&$0$&$-Q$&$0$&$0$&$0$\\
\hline
$T_3$&$0$&$0$&$0$&$-3T_3$&$T_3$&$0$&$0$&$0$&$0$&$0$&$0$\\
\hline
$Q$&$0$&$0$&$0$&$0$&$-Q$&$0$&$0$&$0$&$0$&$0$&$0$\\
\hline
\end{tabular}
\end{center}

The $x$-integral $F(...a,b,a_1,b_1,a_2,b_2\dots)$ must satisfy equations
\begin{equation}
\begin{array}{llll}
A(F)=0, & B(F)=0, &P_1(F)=0,& P_2(F)=0,\\
\label{system3} \\
 T_1(F)=0, & T_2(F)=0, & Q(F)=0, & T_3(F)=0. \\
\end{array}  
\end{equation}

Solving system (\ref{system3}), we assume that $F_1$ depends on $a,\,b_1,\,a_1,\,b_2,\, a_2,\, b_3, \, a_3,\, b_4,\,a_4$ and $F_2$ depends on $b,\,a,\,b_1,\, a_1,\, b_2, \, a_2,\, b_3,\,a_3,\,b_4$, and find two independent $x$-integrals
$$
\begin{array}{ll}
F_1= & e^{-a+a_1}+e^{-2a_1+2a_2+b_2-b_3}+2e^{-a_1+2a_2-a_3}+e^{-a_2+a_3+b_3-b_4}+\\
     & +e^{2a_2-2a_3-b_2+b_3}+e^{a_1-a_2-b_1+b_2}+e^{a_3-a_4},\\
F_2= & e^{-b+b_1}+3e^{-2a+3a_1-a_2}+3e^{-a_1+3a_2-2a_3}+3e^{-a+a_1+a_2-a_3}+e^{-3a+3a_1+b_1-b_2}+ \\  
     & +3e^{-a+3a_1-2a_2-b_1+b_2}+3e^{a_1-a_3-b_1+b_2}+e^{3a_1-3a_2-2b_1+2b_2}+e^{-3a_1+3a_2+2b_2-2b_3}+\\
     & +3e^{-a+a_2+b_2-b_3}+3e^{-2a_1+3a_2-a_3+b_2-b_3} +2e^{-b_1+2b_2-b_3}+e^{3a_2-3a_3-b_2+b_3}+e^{b_3-b_4}.
\end{array}
$$

Here the Cartan matrix is 
$
N=\left(
\begin{array}{cc}
2&-1\\
-3&2
\end{array} \right).
$
The isomorphism between the subalgebra of the characteristic algebra generated by $\{A,B,P_1,P_2\}$ and the Borel subalgebra of the algebra $G_2$ is established by the same mapping as in the previous example.
\end{example}
{\bf Remark 2}. In order to find the arguments of $F$ one has to order the set of the dynamical variables $\{a_i,\,b_i\}$ in an alternating way as follows 
$$
\dots \,b_{-1},\,a_{-1},\,b,\,a,\,b_1,\,a_1,\, b_2, \, a_2,\, b_3,\,a_3,\,b_4,\,a_4,\,\dots
$$
and then take a segment of the length greater by one than the dimension of the system of equations for $F$. If for one of the integrals
the segment begins with $a$ then for the other one it begins with $b$. 
 
\section{Construction of $n$-integrals for differential-difference systems}

It turns out that due to the special form of system (\ref{sys}) it is reasonable to look for $n$-integrals using the definition. We will search an $n$-integrals which  are in a sense homogeneous polynomials of the derivatives of the variables $a$, $b$. That is they are linear combinations of  monomials 
\begin{equation}
a_{[1]}^{p_1}a_{[2]}^{p_2}\dots a_{[k]}^{p_k}b_{[1]}^{q_1}b_{[2]}^{q_2}\dots b_{[k]}^{q_k}
\end{equation}
for which the sum $\sum_{i=1}^k (ip_i+iq_i)$  is constant. We also note that one can always find independent integrals that are   linear in the  highest derivative present (see \cite{HZhS}). Hence, to find an $n$-integral one has to fix the order of the highest derivative. Then write the general homogeneous, in the above sense, polynomial  $I$ with undetermined coefficients. The coefficients are found using the equality $DI=I$.    
\begin{example} 
For $c=1$  the $n$-integrals of (\ref{sys}) are
$$
\begin{array}{l}
I_1=a_{xx}+ b_{xx}-a_{x}^2+a_{x}b_{x}-b_{x}^2,\\
I_2=a_{[3]}+a_{x}(b_{xx}-2a_{xx})+a_{x}^2b_{x}-a_{x}b_{x}^2.
\end{array}
$$
\end{example}

\begin{example} 
For $c=2$  the $n$-integrals of (\ref{sys}) are
$$
\begin{array}{l}
I_1=2a_{xx}+b_{xx}-2a_{x}^2+2a_{x}b_{x}-b_{xx}^2,\\
I_2=a_{[4]}+a_{x}(b_{[3]}-2a_{[3]})+a_{xx}(4a_{x}b_{x}-2a_{x}^2-b_{x}^2)+\\
\qquad +a_{xx}(b_{xx}-a_{xx})+b_{xx}a_{x}(a_{x}-2b_{x})+a_{x}^4+a_{x}^2b_{x}^2-2a_{x}^2b_{x}.
\end{array}
$$ 
\end{example}

\begin{example} 
For $c=3$  the $n$-integrals of (\ref{sys}) are
$$
\begin{array}{l}
I_1=a_{xx}+\frac{1}{3}b_{xx}-(a_{x})^2+a_{x}b_{x}-\frac{1}{3}b_{x}^2,\\
I_2= a_{[6]}-2a_{[5]}a_{x}+ b_{[5}]a_{x}+32a_{[4]}a_{x}^2- 30a_{[4]}a_{x}b_{x}+ 11a_{[4]}b_{x}^2-40a_{[4]}a_{xx}- \\
\qquad
-11a_{[4]}b_{xx}+14b_{[4]}a_{x}^2 -15b_{[4]}a_{x}b_{x} +\frac{13}{3}b_{[4]}b_{x}^2 -10b_{[4]}a_{xx} -\frac{13}{3}b_{[4]}b_{xx} +\\
\qquad
+19a_{[3]}^2+\frac{13}{6}b_{[3]}^2 +16a_{[3]}b_{[3]} -36a_{[3]}a_{xx}a_{x} +18a_{[3]}a_{xx}b_{x} +80a_{[3]}b_{xx}a_{x} - \\
\qquad
-45a_{[3]}b_{xx}b_{x} -52b_{[3]}a_{xx}a_{x} +33b_{[3]}a_{xx}b_{x} -5b_{[3]}b_{xx}a_{x} -64a_{[3]}a_{x}^3 + \\
\qquad
+102a_{[3]}a_{x}^2b_{x} -2a_{[3]}a_{x}b_{x}^2 +13a_{[3]}b_{x}^3 +32b_{[3]}a_{x}^3 -58b_{[3]}a_{x}^2b_{x} + 38b_{[3]}a_{x}b_{x}^2-\\
\qquad
 -\frac{26}{3}b_{[3]}b_{x}^3+ 66a_{xx}^3 + \frac{26}{3}b_{xx}^3 -35a_{xx}^2b_{xx} -5a_{xx}b_{xx}^2 + 30a_{xx}^2a_{x}^2- 18a_{xx}^2a_{x}b_{x} -\\
\qquad
 -\frac{11}{2}a_{xx}^2b_{x}^2 -34a_{xx}b_{xx}a_{x}^2 +32a_{xx}b_{xx}a_{x}b_{x}-2a_{xx}b_{xx}b_{x}^2 -2b_{xx}a_{x}b_{x} +6a_{x}a_{x}^4 -\\
\qquad
-24a_{xx}a_{x}^3b_{x} +25a_{xx}a_{x}^2b_{x}^2 - 9a_{xx}a_{x}b_{x}^3 +a_{xx}b_{x}^4 -b_{xx}a_{x}^4+8b_{xx}a_{x}^3b_{x} -8b_{xx}a_{x}^2b_{x}^2 +\\
\qquad
+2b_{xx}a_{x}b_{x}^3 -2a_{x}^6 +6a_{x}^5b_{x} - \frac{13}{2}a_{x}^4b_{x}^2 +3a_{x}^3b_{x}^3 - \frac{1}{2}a_{x}^2b_{x}^4.
\end{array}
$$
 \end{example}
 
 It is remarkable that $n-$integrals for the differential-difference chain (\ref{sys}) given in Examples 4-6 coincide with $y-$integrals for the corresponding exponential type PDE (\ref{cont_system1}) with the same Cartan matrix. Recall that in the continuous case these integrals except the last one in Example 6 were found in \cite{LSSh1982} (see also \cite{HandB}). 

\section{Integrable cutting of constraints and finite systems}

This section dwells upon alternative approach to construct integrable  differential-difference chains of exponential type. The approach is based on finding the integrable boundary conditions for soliton systems \cite{H}, \cite{KhG}.
In what follows we  write functions, that depend on a continuous variable $x$ and two discrete variables $n$, $j$   
as $f(x,n,j)=f^n$ and denote its shifts as $f(x,n,j+k)=f^n_k$.
Consider the  equation
\begin{equation}\label{origin}
u^n_{x}- u^n_{1,x}=e^{u^{n-1}-u^{n}_1}-e^{u^n-u^{n+1}_1} 
\end{equation}
realizing the B\"acklund transformation for the well known Toda lattice. Here the upper index $n$ enumerates the field  variables. In equation (\ref{origin}) the variables $ \dots,u^{-2}, u^{-1}, u^0, u^1, u^2,\dots $ are considered as dynamical ones. Below we study finite field reduction of the chain.

Equation (\ref{origin}) is the compatibility condition of the following pair of linear equations
\begin{equation}\label{lax1}
\begin{array}{l}
\phi^n_{x}=\phi^{n+1}-u^n_{x}\phi^n,\\
\phi^{n}_1=e^{u^{n-1}-u^{n}_1}\phi^{n-1}+\phi^n,
\end{array}
\end{equation}
which can be written as a discrete version of the Lax equation. Indeed, let $U:=D_n-u_x^n$, $V:=e^{u^{n-1}-u_1^n}D_n^{-1}+1$, then (\ref{lax1}) takes the form $\phi_x=U\phi$, $D_j\phi=V\phi$, where $D_n, D_j$ are the shift operators acting due to the rule $D_n\phi^n=\phi^{n+1}$, $D_j\phi=\phi_1$. Now evidently the compatibility condition of the system reads as 
$$
V_x=D_j(U)V-VU.
$$
The latter coincides with the discrete Lax equation. We refer below to (\ref{lax1}) as a Lax pair.

Excluding $\phi^{n-1}$ from a pair of the equations (\ref{lax1}) we obtain a hyperbolic type linear differential-difference equation
\begin{equation}\label{eqn1.1}
\phi^n_{1,x}-\phi^n_x+u^n_{1,x}\phi^n_1-
(u^n_x+ e^{u^{n}-u^{n+1}_1})\phi^{n}=0.
\end{equation}
The equation (\ref{origin}) is invariant with respect to the  transformation defined as $u\to -u$, $x\to -x$, $n\to -n$ and $j\to -j$.
Under this transformation the Lax pair (\ref{lax1})  transforms to a Lax pair
\begin{equation}\label{lax2}
\begin{array}{l}
\psi^n_{x}=-\psi^{n-1}+u^n_{x}\psi^n,\\
\psi^{n}_{-1}=e^{u^{n}_{-1}-u^{n+1}}\psi^{n+1}+\psi^n
\end{array}
\end{equation}
and equation (\ref{eqn1.1}) transforms to the equation
\begin{equation}\label{eqn2.1} 
\psi^n_{1,x}-\psi^n_{x}+[-u^n_{1,x}-e^{u^{n-1}-u^{n}_1}]\psi^{n}_1+u^n_{x}\psi^{n}=0.
\end{equation}

Thus we have two families (\ref{eqn1.1}), (\ref{eqn2.1}) of linear differential-difference equations enumerated by $n$. We would like to know when  an equation from one family can be related, by a linear transformation, to  an equation from the other family. To this end we evaluate the Laplace invariants of these equations (see \cite{AdlerStartsev}). 

Recall that the Laplace invariants of the equation \cite{AdlerStartsev}:
\begin{equation}
y_x(j+1)+a(j)y_x(j)+b(j)y(j+1)+c(j)y(j)=0
\end{equation} 
are given by $K_1=\displaystyle{\frac{c(j)-a(j)b(j-1)}{a(j)}}$ and $K_2=\displaystyle{\frac{c(j)-a(j)b(j)-a_x(j)}{a(j)}}$. By virtue of these formulas the invariants $K_{1\phi}, K_{2\phi}$ and $K_{1\psi},K_{2\psi}$ of equations (\ref{eqn1.1}), (\ref{eqn2.1}) are, respectively,
\begin{eqnarray}
K_{1\phi}&=&e^{u^{n}-u^{n+1}_{1}}, K_{2\phi}=e^{u^{n-1}-u^{n}_1},\\
K_{1\psi}&=&e^{u^{n-1}_{-1}-u^{n}},K_{2\psi}=e^{u^{n}-u^{n+1}_1}.
\end{eqnarray}

It is known that two linear hyperbolic type equations are related to one another by a linear transformation only if their  corresponding Laplace invariants are equal. Evidently in generic case coincidence of the Laplace invariants generates two constraints on the field variables $u^n(x,j)$. Only for some special cases it gives only one constraint. We are interested in such special cases. For instance pair of equations $K_{1\phi}(n,j)=K_{1\psi}(n+1,j+1)$, $K_{2\phi}(n,j)=K_{2\psi}(n+1,j+1)$ is equivalent to the constraint
\begin{equation}\label{bound_cond_original}
u^{n-1}_{-1}-u^{n}=u^{n+1}-u^{n+2}_1, \quad \forall j \in \bf Z,
\end{equation}
which is interpreted as a cutting off boundary condition for the chain (\ref{origin}).
For simplicity we put $n=0$, so the boundary condition becomes 
\begin{equation}\label{bound_cond}
u^{-1}_{-1}-u^{0}=u^{1}-u^{2}_1.
\end{equation}
Following \cite{H} we can construct a Lax pair of the corresponding reduction.
Under the boundary condition (\ref{bound_cond}) we have equality of the invariants
\begin{equation}
\begin{array}{c}
K_{1,\phi}(1,j)=K_{1,\psi}(0,j),\\
K_{2,\phi}(1,j)=K_{2,\psi}(0,j)\\
\end{array}
\end{equation}
and
\begin{equation}
\begin{array}{c}
K_{1,\psi}(1,j)=K_{1,\phi}(0,j-1),\\
K_{2,\psi}(1,j)=K_{2,\phi}(0,j-1)\\
\end{array}
\end{equation}
and  we can relate the eigenfunctions 
\begin{equation}\label{AB}
\phi^0=A(x,j)\psi^1_1 \qquad \mbox{and} \qquad \psi^0=B(x,j)\phi^1.
\end{equation}
Hence, we have from (\ref{lax1}), (\ref{lax2}), (\ref{AB})
\begin{equation}\label{dynam1}
\begin{array}{l}
\phi^1_1=e^{u^0-u^1_1}A\psi^1_1+\phi^1,\\
\psi^1_x=-B\phi^1+u^1_x\psi^1.
\end{array}
\end{equation}

We study the finite reductions of the chain (\ref{origin}) on a finite interval $N_L\leq n \leq N_R$. The reduction is obtained by imposing the boundary conditions at the left end-point $n=N_L$
\begin{equation}\label{L}
u^{N_L-1}=u^{N_L}_1+u^{N_L+1}_1-u^{N_L+2}_2
\end{equation}
and respectively at the right end-point $n=N_R$
\begin{equation}\label{R}
u^{N_R+1}=u^{N_R}_{-1}+u^{N_R-1}_{-1}-u^{N_R-2}_{-2}.
\end{equation}

First we concentrate on the left end-point. Due to the reasonings above the eigenfunctions should satisfy the following gluing conditions
\begin{equation}\label{glu}
\phi^{N_L}=A\psi^{N_L+1}_1 \qquad \mbox{and} \qquad \psi^{N_L}=B\phi^{N_L+1}.
\end{equation}

These conditions allow one to close the Lax equations at the left end-point
\begin{eqnarray}\label{dynam_phi1}
\phi^{N_L+1}_1&=&Ae^{u^{N_L}-u^{N_L+1}_1}\psi^{N_L+1}_1+\phi^{N_L+1},\nonumber\\
\psi^{N_L+1}_{x}&=&-B\phi^{N_L+1}+u^{N_L+1}_{x}\psi^{N_L+1}.
\end{eqnarray}

The compatibility conditions of these equations with their counterparts (\ref{lax1}), (\ref{lax2}):
\begin{eqnarray}
\phi^{N_L+1}_x&=&\phi^{N_L+2}-u^{N_L+1}_x\phi^{N_L+1},\nonumber\\
\psi^{N_L+1}_{-1}&=&e^{u^{N_L+1}_{-1}-u^{N_L+2}}\psi^{N_L+2}+\psi^{N_L+1}
\end{eqnarray}
yield an overdetermined system of equations for $A,B$:
\begin{equation}\label{A}
\begin{array}{l}
A(j+1)=A(j),\\
A_x(j)+\left(u^{N_L}_x+u^{N_L+1}_{1,x}+e^{u^{N_L}-u^{N_L+1}_1}\right)A(j)=0
\end{array}
\end{equation}
and
\begin{equation}\label{B}
\begin{array}{l}
B(j+1)=B(j),\\
B_x(j)=\left(u^{N_L}_x+u^{N_L+1}_x+e^{u^{N_L+1}-u^{N_L+2}_1}\right)B(j).
\end{array}
\end{equation}
The compatibility of the above equations is guaranteed by the boundary condition (\ref{bound_cond}). The functions 
$A$ and $B$ are determined up to a constant and can be chosen so that $A\cdot B=-1$. So, we put
\begin{equation}
B=\lambda \bar B \quad \mbox{and} \quad A=-(\lambda \bar B)^{-1} 
\end{equation}
where $\bar B$ is a particular solution of (\ref{B}), $\lambda$ is a constant parameter.

Put $\rho^{N}(x,j)=\psi^N(x,j+1)$ for the convenience and find
\begin{eqnarray}\label{s_phi}
\phi^{N_L+1}_1&=&Ae^{u^{N_L}-u^{N_L+1}_1}\rho^{N_L+1}+\phi^{N_L+1},\quad 
\phi^{N_L+1}_{x}=\phi^{N_L+2}-u^{N_L+1}_{x}\phi^{N_L+1},\nonumber\\
\phi^{n}_1&=&e^{u^{n-1}-u^n_1}\phi^{n-1}+\phi^n, \quad
\phi^n_x=\phi^{n+1}-u^n_x\phi^n \quad \mbox{for}\quad  N_L+2\leq n\leq N_R-2
\end{eqnarray}
and
\begin{eqnarray}\label{s_rho}
\rho^{N_L+1}_{-1}&=&e^{u^{N_L+1}-u^{N_L+2}_1}\rho^{N_L+2}+\rho^{N_L+1},\quad
\rho^{N_L+1}_x=-B\phi^{N_L+1}+(u^{N_L+1}_{1,x}+e^{u^{N_L}-u^{N_L+1}_1})\rho^{N_L+1},\nonumber\\
\rho^n_{-1}&=&e^{u^n-u^{n+1}_1}\rho^{n+1}+\rho^n,\quad
\rho^n_x=-\rho^{n-1}+u^n_{1,x}\rho^n, \quad \mbox{for} \quad N_L+2\leq n\leq N_R-2.
\end{eqnarray}

To derive similar equations at the point $N_R$ we use the right end-point constraint (\ref{R}), for which we have 
\begin{equation}
\psi^{N_R}_1=\hat{A}\phi^{N_R-1}, \quad \phi^{N_R}=\hat{B}\psi^{N_R-1}\nonumber
\end{equation}
where 
\begin{equation}\label{hat_A}
\begin{array}{l}
\hat{A}(j+1)=\hat{A}(j),\\
\hat{A}_x(j)=\left(u^{N_R}_x+u^{N_R-1}_{-1,x}+e^{u^{N_R-1}_{-1}-u^{N_R}}\right)\hat{A}(j)=0
\end{array}
\end{equation}
and
\begin{equation}\label{hat_B}
\begin{array}{l}
\hat{B}(j+1)=\hat{B}(j),\\
\hat{B}_x(j)=-\left(u^{N_R}_x+u^{N_R-1}_x+e^{u^{N_R-2}_{-1}-u^{N_R-1}}\right)\hat{B}(j).
\end{array}
\end{equation}
The functions $\hat{A}$ and $\hat{B}$ are determined up to a constant and can be chosen so that $\hat{A}\cdot \hat{B}=-1$.
We put $\hat{A}=\lambda \hat{A}_1$ and $\hat{B}=-(\lambda \hat{A}_1)^{-1}$, where $\hat{A}_1$ is a particular solution of  (\ref{hat_A}). 

And finally, we find in addition to (\ref{s_phi}), (\ref{s_rho}) a part of Lax equations corresponding to the right end-point
\begin{eqnarray}\label{sR}
\phi^{N_R-1}_1&=&e^{u^{N_R-2}-u^{N_R-1}_1}\phi^{N_R-2}+\phi^{N_R-1},\quad
\phi^{N_R-1}_x=\hat{B}\rho^{N_R-1}-(u^{N_R-1}_x+e^{u^{N_R-3}_{-1}-u^{N_R-2}})\phi^{N_R-1}, \nonumber\\
\rho^{N_R-1}_{-1}&=&e^{u^{N_R-1}-u^{N_R}_1}\hat{A}\phi^{N_R-1}+\rho^{N_R-1},\quad
\rho^{N_R-1}_x=-\rho^{N_R-2}+u^{N_R-1}_{1,x}\rho^{N_R-1}.
\end{eqnarray} 

\begin{theorem}
Chain (\ref{origin}) reduced to the finite interval $[N_L,N_R]$ with the boundary condition (\ref{L}), (\ref{R}) is the compatibility condition of the overdetermined system of linear equations (\ref{A}), (\ref{B}), (\ref{s_phi})-(\ref{sR}).
\end{theorem}
{\bf Proof.} We check the compatibility conditions of the equations (\ref{s_phi})-(\ref{sR}) for $\phi^n,\rho^n,\quad 
N_L+1\leq n\leq N_R-1$.
It is easy to see that the compatibility condition for $\phi^{N_L+1}$ or for $\rho^{N_L+1}$ leads to 
$$
u_x^{N_L+1}-u_{1,x}^{N_L+1}=e^{u^{N_L}-u^{N_L+1}_1}-e^{u^{N_L+1}-u^{N_L+2}_1}. 
$$
Similarly, the compatibility condition for $\phi^{N_R-1}$ or for $\rho^{N_R-1}$ leads to 
$$
u_x^{N_R-1}-u_{1,x}^{N_R-1}=e^{u^{N_R-2}-u^{N_R-1}_1}-e^{u^{N_R-1}-u^{N_R}_1}.
$$
The compatibility condition for $\phi^n$ or for $\rho^n$ where $N_L+2\leq n\leq N_R-2$ leads to 
$$
u_x^n-u_{1,x}^n=e^{u^{n-1}-u^{n}_1}-e^{u^{n}-u^{n+1}_1},\quad N_L+2\leq n\leq N_R-2.
$$
The compatibility condition for $A$ or for $B$ leads to 
$$
u_x^{N_L}-u_{1,x}^{N_L}=e^{u^{N_L+1}_1-u^{N_l+2}_2}-e^{u^{N_L}-u^{N_L+1}_1}.
$$
And finally, the compatibility condition for $\hat{A}$ or for $\hat{B}$ leads to 
$$
u_x^{N_R}-u_{1,x}^{N_R}=e^{u^{N_R-1}-u^{N_R}_1}-e^{u^{N_R-2}_{-1}-u^{N_R-1}}.\\
 \Box
$$ 

In addition to boundary condition (\ref{bound_cond}) we  introduce  the degenerate boundary condition $e^{-u^{N+1}}=0$, for some $N$ and for all integer $j$. The degenerate boundary condition implies that the corresponding eigenfunction is zero, $\phi^{N+1}=0$.

The chain (\ref{origin}) can be related to the  chain
\begin{equation}\label{nc}
r^n_x-r^n_{1,x}=e^{r^{n-1}-r^n-r^n_1+r^{n+1}_1},
\end{equation}
which was considered in the paper \cite{AdlerStartsev}, by the transformation
\begin{equation}\label{u_r}
u^n=r^{n-1}-r^n.
\end{equation}
The correspondence between boundary conditions is as follows.
The condition $u^{-1}=u^0_1+u^1_1-u^2_2$ leads to the condition 
\begin{equation}\label{***}
r^{-1}=r^1_1,
\end{equation}
the condition $u^{N+1}_1=-u^{N-2}_{-1}+u^{N-1}+u^{N}$ leads to the condition 
\begin{equation}\label{*****}
r^{N}_1=r^{N-2},
\end{equation}
and the degenerate conditions $e^{u^0}=0$ and $e^{-u^{N+1}}=0$ lead respectively to the conditions 
\begin{equation}\label{******}
r^0=0,
\end{equation}
\begin{equation}\label{****}
r^{N+1}=0.
\end{equation}

\subsection[exCn]{Integrable systems corresponding to the Cartan matrices $C_N$}

Now we consider reductions of chain (\ref{origin}). We impose condition (\ref{bound_cond}) for $n=0$ and the degenerate boundary condition $e^{-u^{N+1}}=0$ for $n=N+1$. The resulting reduction is as follows 
\begin{equation}\label{chain2_u}
\begin{array}{lll}
u^0_{x}- u^0_{1,x}&=&e^{u^1_1-u^2_2}-e^{u^0-u^1_1}, \\
u^1_{x}- u^1_{1,x}&=&e^{u^{0}-u^{1}_1}-e^{u^1-u^{2}_1},\\
 & \dots & \\
u^{N-1}_{x}- u^{N-1}_{1,x}&=&e^{u^{N-2}-u^{N-1}_1}-e^{u^{N-1}-u^{N}_1}, \\
u^N_{x}- u^N_{1,x}&=&e^{u^{N-1}-u^{N}_1}. 
\end{array}
\end{equation}

The above reduction admits a Lax pair.
Indeed, let us consider vectors $\phi=(\phi^1,\phi^2, \dots,\phi^N), \rho=(\rho^1,\rho^2, \dots,\rho^N)$ and 
denote $\gamma^n=e^{u^n-u^{n+1}_1}$. We introduce
$N\times N$ matrices
$$
u_{11}=\left(
\begin{array}{cccccc}
 1           & 0     & 0 & 0            & \dots          & 0       \\
 \gamma^1    & 1     & 0 & 0            & \dots          & 0       \\
             &       &   & \dots        &                &         \\ 
 0           & \dots & 0 & \gamma^{N-2} & 1              & 0       \\        
 0           & \dots & 0 & 0            & \gamma^{N-1}& 1        \\  
\end{array}
\right),\,
u_{12}=\left(
\begin{array}{cccc}
A\gamma^0    & 0     &  \dots & 0 \\
0            & 0     &  \dots & 0 \\
             & \dots &        &   \\ 
0            & 0     &  \dots & 0 \\
\end{array}
\right),
$$
$$
u_{22}=\left( \begin{array}{ccccc}
1&\gamma^1&0       &\dots&0\\
0&1       &\gamma^2&\dots&0\\
 &        &\dots   &     & \\
0&\dots   &0       &1    &\gamma^{N-1}\\
0&\dots   &0       &0    &1
\end{array}
\right),\quad
v_{11}=\left(
\begin{array}{cccccc}
-u^1_{x}& 1      & 0       & 0     & \dots     & 0       \\
 0      &-u^2_{x}& 1       & 0     & \dots     & 0       \\
        &        &         & \dots &           &         \\ 
0       & \dots  &  0      & 0     &-u^{N-1}_x & 1       \\        
0       & 0      & \dots   & 0     & 0         &-u^N_{x} \\                        
\end{array}
\right),
$$
$$
v_{21}=\left( \begin{array}{cccc}
-B&0&\dots&0\\
0 &0&\dots&0\\
0 &0&\dots&0\\
0 &0&\dots&0\\
\end{array}
\right), \quad
v_{22}=\left( \begin{array}{ccccc}
\gamma^0+u^1_{1,x}      & 0        & 0       & \dots        & 0        \\
-1                      & u^2_{1,x}& 0       & \dots        & 0        \\
 0                      & -1       &u^3_{1,x}& \dots        & 0        \\
                        &          & \dots   &              &          \\         
 0                      &   \dots  & 0       & -1           & u^N_x(j) \\                        
\end{array}
\right),
$$
where $A,B$ are particular solutions of (\ref{A}), (\ref{B}).

The following lemma gives the Lax pair for system (\ref{chain2_u}).
\begin{lemma}
The compatibility conditions for the equations (\ref{A}), (\ref{B}),
\begin{equation}
\left(
\begin{array}{cc}
DE&0\\
0&D^{-1}E
\end{array}
\right)
\left(\begin{array}{c}
\phi\\ \rho 
\end{array}
\right)=
\left(
\begin{array}{cc}
u_{11}&u_{12}\\
u_{21}&u_{22}
\end{array}
\right)
\left(\begin{array}{c}
\phi\\ \rho 
\end{array}
\right),\quad
\left(
\begin{array}{c}
\phi_x\\
\rho_x
\end{array}
\right)=
\left(
\begin{array}{cc}
v_{11}&v_{12}\\
v_{21}&v_{22}
\end{array}
\right)
\left(\begin{array}{c}
\phi\\ \rho 
\end{array}
\right),
\end{equation}
where $u_{21}, v_{12}$ are matrices of dimension $N$ with zero entries , $E$ is the unity matrix of dimension $N$, lead to the system (\ref{chain2_u}).
\end{lemma}

If we impose boundary conditions $r^{-1}=r^1_1, \quad r^{N+1}=0$ we obtain the following reduction of chain (\ref{nc})
\begin{equation}
\begin{array}{lll}
r^0_{x}-r^0_{1,x} & = &  e^{-r^0-r^0_1+2r^1_1},\\
r^1_{x}-r^1_{1,x} & = & e^{r^{0}-r^1-r^1_1+r^{2}_1},\\
 & \dots & \\
r^{N-1}_x-r^{N-1}_{1,x} & = & e^{r^{N-2}-r^{N-1}-r^{N-1}_1+r^{N}_1},\\
r^N_{x}-r^N_{1,x}& = & e^{r^{N-1}- r^N-r^N_1},\\
\end{array}
\end{equation}
which is related to (\ref{chain2_u}). This system corresponds to the Cartan matrices $C_{N+1}$ ($B_{N+1}$).

Below we use some manipulations to reduce it to the form (\ref{chain2_u}). First rewrite the system as follows
\begin{equation}
\begin{array}{lll}
r^{-1}&=&r_1^1,\\
r^0_{x}-r^0_{1,x} & = &  e^{-r^0-r^0_1+2r^1_1},\\
r^1_{x}-r^1_{1,x} & = & e^{r^{0}-r^1-r^1_1+r^{2}_1},\\
 & \dots & \\
r^{N-1}_x-r^{N-1}_{1,x} & = & e^{r^{N-2}-r^{N-1}-r^{N-1}_1+r^{N}_1},\\
r^N_{x}-r^N_{1,x}& = & e^{r^{N-1}- r^N-r^N_1},\\
r^{N+1}&=&0.
\end{array}
\end{equation}

Since $r^{-1}=r^1_1$ we have
$r^{-1}_{x}-r^{-1}_{1,x} = e^{r^0_1-r^1_1-r^1_2+r^2_2}.$
Since $r^{N+1}=0$ we have $r^{N+1}_{x}-r^{N+1}_{1,x}=0.$
Thus we can extend the above system to 
\begin{equation}\label{obr3r+gr_usl}
\begin{array}{lll}
r^{-1}_{x}-r^{-1}_{1,x} & = & e^{r^0_1-r^1_1-r^1_2+r^2_2},\\
r^0_{x}-r^0_{1,x} & = &  e^{-r^0-r^0_1+2r^1_1},\\
r^1_{x}-r^1_{1,x} & = & e^{r^{0}-r^1-r^1_1+r^{2}_1},\\
 & \dots & \\
r^{N-1}_x-r^{N-1}_{1,x} & = & e^{r^{N-2}-r^{N-1}-r^{N-1}_1+r^{N}_1},\\
r^N_{x}-r^N_{1,x}& = & e^{r^{N-1}- r^N-r^N_1},\\
r^{N+1}_{x}-r^{N+1}_{1,x}&=&0.\\
\end{array}
\end{equation}
We subtract the second equation from the first one, the third equation from the second one, and so on.
Then introducing variables (\ref{u_r}) for $n=0,1,2,\dots,N$ we arrive at system (\ref{chain2_u}).

\subsection[exAn] {Integrable systems corresponding to the Cartan matrices $A_N$}

We can impose the degenerate boundary conditions at both ends of the chain (\ref{origin}). 
We put $e^{u^0}=0$ and $e^{-u^{N+1}}=0$ that gives the following reduction   
\begin{equation}\label{chain1_u}
\begin{array}{lll}
u^1_{x}- u^1_{1,x}&=&-e^{u^1-u^2_1}, \\
u^2_{x}- u^2_{1,x}&=&e^{u^{1}-u^{2}_1}-e^{u^2-u^{3}_1},\\
 & \dots & \\
u^{N-1}_{x}- u^{N-1}_{1,x}&=&e^{u^{N-2}-u^{N-1}_1}-e^{u^{N-1}-u^{N}_1}, \\
u^N_{x}- u^N_{1,x}&=&e^{u^{N-1}-u^{N}_1}. 
\end{array}
\end{equation}

The system (\ref{chain1_u}) admits a Lax pair that can be obtained from the Lax pair (\ref{lax1}). The boundary conditions imply that the corresponding eigenfunctions are zero, $\phi^0=0$ and $\phi^{N+1}=0$. Consider vectors $\phi=(\phi^1,\phi^2,\dots,\phi^N)$. We introduce
$N\times N$ matrices
\begin{equation}
R=\left(
\begin{array}{cccccc}
1             & 0           & 0     & 0               & \dots           & 0 \\
\gamma^{1} & 1           & 0     & 0               & \dots           & 0 \\
0             & \gamma^2 & 1     & 0               & \dots           & 0 \\       
              &             & \dots &                 &                 &   \\
0             & 0           & \dots & \gamma^{N-2} & 1               & 0 \\        
0             & 0           & \dots & 0               & \gamma^{N-1} & 1  \\                        
\end{array}
\right) \quad 
\mbox{and} \quad 
S=\left(
\begin{array}{cccccc}
-u^1_{x}& 1      & 0       & 0     & \dots     & 0       \\
 0      &-u^2_{x}& 1       & 0     & \dots     & 0       \\
        &        &         & \dots &           &         \\ 
0       & \dots  &  0      & 0     &-u^{N-1}_x & 1       \\        
0       & 0      & \dots   & 0     & 0         &-u^N_{x} \\                        
\end{array}
\right).\nonumber
\end{equation}

\begin{lemma}
The compatibility conditions for the equations 
\begin{equation}
\phi_1=R\phi, \quad \phi_x=S\phi  
\end{equation}
lead to the system (\ref{chain1_u}).
\end{lemma}

Imposing boundary conditions (\ref{******}) and (\ref{****}) we get the following chain related with the chain (\ref{chain1_u})
\begin{equation}\label{system_A}
\begin{array}{lll}
r^1_{x}-r^1_{1,x} &= & e^{-r^1-r^1_1+r^2_1},\\
r^2_{x}-r^2_{1,x} & = & e^{r^{1}-r^2-r^2_1+r^3_1},\\
 & \dots & \\
r^{N-1}_x-r^{N-1}_{1,x} & = & e^{r^{N-2}-r^{N-1}-r^{N-1}_1+r^{N}_1},\\
r^N_{x}-r^N_{1,x}& = & e^{r^{N-1}- r^N-r^N_1}.\\
\end{array}
\end{equation}
The above system corresponds to the Cartan matrices $A_N$.

The system (\ref{system_A}) can be reduced to the system
\begin{equation}\label{J}
\begin{array}{lll}
v^1_{x}-v^1_{1,x} &= & -e^{v^1}-e^{v^1_1}+e^{v^2_1},\\
v^2_{x}-v^2_{1,x} &= & e^{v^{1}}-e^{v^2}-e^{v^2_1}+e^{v^3_1},\\
 & \dots & \\
v^{N-1}_x-v^{N-1}_{1,x} & = & e^{v^{N-2}}-e^{v^{N-1}}-e^{v^{N-1}_1}+e^{v^{N}_1},\\
v^N_{x}-v^N_{1,x}& = & e^{v^{N-1}}- e^{v^N}-e^{v^N_1},\\
\end{array}
\end{equation}
called two-dimensional Toda molecule equation, found in paper  \cite{Hirota}, by the transformation
\begin{equation}\label{r_v}
e^{v^n}=r^n_x-r^n_{1,x}.
\end{equation}
The correspondence between boundary conditions is as follows.
The conditions $r^0=0$, $r^{N+1}=0$ lead respectively to the conditions $e^{v^0}=0$, $e^{v^{N+1}}=0$.

\subsection[exDn] {Integrable systems corresponding to the matrices $D^{(2)}_{N}$}

We can impose  boundary conditions (\ref{bound_cond_original}) at both ends of the chain (\ref{origin}). 
That is for $n=0$ we have (\ref{bound_cond}) and for $n=N$ we have 
\begin{equation}\label{bound_cond_N}
u^{N-2}_{-1}-u^{N-1}=u^{N}-u^{N+1}_1.
\end{equation}
This leads to the following reduction  
\begin{equation}\label{chain3_u}
\begin{array}{lll}
u^0_{x}- u^0_{1,x}&=&e^{u^1_1-u^2_2}-e^{u^0-u^1_1}, \\
u^1_{x}- u^1_{1,x}&=&e^{u^{0}-u^{1}_1}-e^{u^1-u^{2}_1},\\
 & \dots & \\
u^{N-1}_{x}- u^{N-1}_{1,x}&=&e^{u^{N-2}-u^{N-1}_1}-e^{u^{N-1}-u^{N}_1}, \\
u^N_{x}- u^N_{1,x}&=&e^{u^{N-1}-u^{N}_1}-e^{u^{N-2}_{-1}-u^{N-1}}. 
\end{array}
\end{equation}

Imposing boundary conditions (\ref{***}) and (\ref{*****}) we get the following chain:
\begin{equation}
\begin{array}{lll}
r^0_{x}-r^0_{1,x} & = &  e^{-r^0-r^0_1+2r^1_1},\\
r^1_{x}-r^1_{1,x} & = & e^{r^0-r^1-r^1_1+r^2_1},\\
 & \dots & \\
r^{N-1}_x-r^{N-1}_{1,x} & = & e^{r^{N-2}-r^{N-1}-r^{N-1}_1+r^{N}_1},\\
r^N_{x}-r^N_{1,x}& = & e^{2r^{N-1}- r^N-r^N_1},\\
\end{array}
\end{equation}
which is related to (\ref{chain3_u}). This system corresponds to the Cartan matrices $D^{(2)}_{N+1}$.\\

Consider vectors $\phi=(\phi^1,\phi^2, \dots,\phi^{N-1}), \rho=(\rho^1,\rho^2, \dots,\rho^{N-1})$. We introduce
$N-1\times N-1$ matrices
$$
u_{11}=\left(
\begin{array}{cccccc}
 1           & 0     & 0 & 0               & \dots          & 0       \\
 \gamma^1    & 1     & 0 & 0               & \dots          & 0       \\
             &       &   & \dots           &                &         \\ 
 0           & \dots & 0 & \gamma^{N-3}    & 1              & 0       \\        
 0           & \dots & 0 & 0               & \gamma^{N-2}   & 1        \\                        
\end{array}
\right),\,
u_{12}=\left(
\begin{array}{cccc}
A\gamma^0 & 0     &  \dots & 0 \\
0         & 0     &  \dots & 0 \\
          & \dots &        &   \\ 
0         & 0     &  \dots & 0 \\
\end{array}
\right),
$$
$$
u_{21}=\left( \begin{array}{cccc}
0&0&\dots&0\\
0&0&\dots&0\\
 & &\dots& \\
0&0&\dots&\hat{A}\gamma^{N-1}
\end{array}
\right),\quad
u_{22}=\left( \begin{array}{ccccc}
1&\gamma^1&0       &\dots&0\\
0&1       &\gamma^2&\dots&0\\
 &        &\dots   &     & \\
0&\dots   &0       &1    &\gamma^{N-2}\\
0&\dots   &0       &0    &1
\end{array}
\right),
$$
$$
v_{11}=\left(
\begin{array}{cccccc}
-u^1_{x}& 1      & 0       & 0     & \dots     & 0       \\
 0      &-u^2_{x}& 1       & 0     & \dots     & 0       \\
        &        &         & \dots &           &         \\ 
0       & \dots  &  0      & 0     &-u^{N-2}_x & 1       \\        
0       & 0      & \dots   & 0     & 0         &-u^{N-1}_{x}-\gamma^{N-3}_{-1} \\ 
\end{array}
\right),
v_{12}=\left( \begin{array}{cccc}
0&0&\dots&0\\
0&0&\dots&0\\
 & &\dots& \\
0&0&\dots&\hat{B}
\end{array}
\right),
$$
$$
v_{21}=\left( \begin{array}{cccc}
-B&0&\dots&0\\
0 &0&\dots&0\\
  &0&\dots&0\\
0 &0&\dots&0
\end{array}
\right), \quad
v_{22}=\left( \begin{array}{ccccc}
\gamma^1_1+u^1_{1,x}    & 0        & 0       & \dots        & 0        \\
-1                      & u^2_{1,x}& 0       & \dots        & 0        \\
 0                      & -1       &u^3_{1,x}& \dots        & 0        \\
                        &          & \dots   &              &          \\         
 0                      &   \dots  & 0       & -1           & u^{N-1}_x(j)                         
\end{array}
\right),
$$
where $A,B,\hat{A},\hat{B}$ are particular solutions of (\ref{A}), (\ref{B}), (\ref{hat_A}), (\ref{hat_B}).

The following lemma gives the Lax pair for system (\ref{chain3_u}).
\begin{lemma}
The compatibility conditions for the equations (\ref{A}), (\ref{B}), (\ref{hat_A}), (\ref{hat_B}),
\begin{equation}
\left(
\begin{array}{cc}
DE&0\\
0&D^{-1}E
\end{array}
\right)
\left(\begin{array}{c}
\phi\\ \rho 
\end{array}
\right)=
\left(
\begin{array}{cc}
u_{11}&u_{12}\\
u_{21}&u_{22}
\end{array}
\right)
\left(\begin{array}{c}
\phi\\ \rho 
\end{array}
\right),\quad
\left(
\begin{array}{c}
\phi_x\\
\rho_x
\end{array}
\right)=
\left(
\begin{array}{cc}
v_{11}&v_{12}\\
v_{21}&v_{22}
\end{array}
\right)
\left(\begin{array}{c}
\phi\\ \rho 
\end{array}
\right)
\end{equation}
where $E$ is the unity matrix of dimension $N-1$, lead to the system (\ref{chain3_u}).
\end{lemma}

\begin{example}
Put $N=3$.
That is  we have the chain 
\begin{equation}
u^n_{x}- u^n_{1,x}=e^{u^{n-1}-u^{n}_1}-e^{u^n-u^{n+1}_1}, \quad n=0,1,2,3,
\end{equation}
with boundary conditions $u^{-1}-u^0_1=u^1_1-u^2_2 \quad \mbox{and} \quad 
u^1_{-1}-u^2=u^3-u^4_1.$
So we have the system
\begin{equation}\label{chain_4}
\begin{array}{lll}
u^0_{x}- u^0_{1,x}&=&e^{u^1_1-u^2_2}-e^{u^0-u^1_1}, \\
u^1_{x}- u^1_{1,x}&=&e^{u^2_1-u^3_2}-e^{u^1-u^2_1}, \\
u^2_{x}- u^2_{1,x}&=&e^{u^1-u^2_1}-e^{u^2-u^3_1},\\
u^3_{x}- u^3_{1,x}&=&e^{u^2-u^3_1}-e^{u^1_{-1}-u^2}.\\ 
\end{array}
\end{equation}
Let us give the Lax pair for the system (\ref{chain_4}) in an explicit form:
the compatibility conditions for the equations (\ref{A}), (\ref{B}), (\ref{hat_A}), (\ref{hat_B}) and
\begin{equation}
\left(
\begin{array}{cc}
DE&0\\
0&D^{-1}E
\end{array}
\right)
\left(\begin{array}{c}
\phi\\ \rho 
\end{array}
\right)=
\left(
\begin{array}{cc}
u_{11}&u_{12}\\
u_{21}&u_{22}
\end{array}
\right)
\left(\begin{array}{c}
\phi\\ \rho 
\end{array}
\right),\quad
\left(
\begin{array}{c}
\phi_x\\
\rho_x
\end{array}
\right)=
\left(
\begin{array}{cc}
v_{11}&v_{12}\\
v_{21}&v_{22}
\end{array}
\right)
\left(\begin{array}{c}
\phi\\ \rho 
\end{array}
\right)
\end{equation}
where
$\phi=(\phi^1,\phi^2)^T, \quad \rho=(\rho^1,\rho^2)^T,$
$$
u_{11}=\left(\begin{array}{cc}
1&0\\
\gamma^1&1\\
\end{array}\right), \quad 
u_{12}=\left(\begin{array}{cc}
A\gamma^0&0\\
0&0\\
\end{array}\right), \quad
u_{21}=\left(\begin{array}{cc}
0&0\\
0&\hat{A}\gamma^2
\end{array}\right), \quad
u_{22}=\left(\begin{array}{cc}
1&\gamma^1\\
0&1\\
\end{array}\right),
$$
$$
v_{11}=\left(\begin{array}{cc}
-u_x^1&1\\
0&-u_x^2-\gamma^0_{-1}
\end{array}\right), \quad 
v_{12}=\left(\begin{array}{cc}
0&0\\
0&\hat{B}
\end{array}\right),
$$
$$
v_{21}=\left(\begin{array}{cc}
-B&0\\
0&0\\
\end{array}\right),\quad
v_{22}=\left(\begin{array}{cc}
u_{1,x}^1+\gamma^1&0\\
-1&u_{1,x}^2
\end{array}\right).
$$

If we impose boundary conditions $r^{-1}=r^1_1, \quad r^{1}=r^{3}_1$ we obtain the following reduction of chain (\ref{nc})
\begin{equation}\label{obr3r}
\begin{array}{lll}
r^0_{x}-r^0_{1,x} & = &  e^{-r^0-r^0_1+2r^1_1},\\
r^1_{x}-r^1_{1,x} & = & e^{r^0-r^1-r^1_1+r^2_1},\\
r^2_{x}-r^2_{1,x} & = & e^{2r^1-r^2-r^2_1}.\\
\end{array}
\end{equation}
The above system corresponds to the Cartan matrix of $D^{(2)}_3$. Below we use some manipulations to reduce it to the form (\ref{chain3_u}). First rewrite the system as follows
\begin{equation}
\begin{array}{lll}
r^{-1} &= & r^1_1 \\
r^0_{x}-r^0_{1,x} & = &  e^{r^{-1}-r^0-r^0_1+r^1_1},\\
r^1_{x}-r^1_{1,x} & = &  e^{r^0-r^1-r^1_1+r^2_1},\\
r^2_{x}-r^2_{1,x} & = &  e^{r^1-r^2-r^2_1+r^3_1},\\
r^{1}& = & r^3_1.\\
\end{array}
\end{equation}

Since $r^{-1}=r^1_1$ we have
$r^{-1}_{x}-r^{-1}_{1,x} = e^{r^0_1-r^1_1-r^1_2+r^2_2}.$
Since $r^{3}=r^1_{-1}$ we have $r^3_{x}-r^3_{1,x}=e^{r^0_{-1}-r^1_{-1}-r^1+r^2}.$
Thus we can extend the above system to 
\begin{equation}
\begin{array}{lll}
r^{-1}_{x}-r^{-1}_{1,x} & = & e^{r^0_1-r^1_1-r^1_2+r^2_2},\\
r^0_{x}-r^0_{1,x}& = & e^{r^{-1}-r^0-r^0_1+r^1_1},\\
r^1_{x}-r^1_{1,x} & = &  e^{r^0-r^1-r^1_1+r^2_1},\\
r^2_{x}-r^2_{1,x} & = &  e^{r^1-r^2-r^2_1+r^3_1},\\
r^3_{x}-r^3_{1,x} & = & e^{r^0_{-1}-r^1_{-1}-r^1+r^2}.\\
\end{array}
\end{equation}
We subtract the second equation from the first one, the third equation from the second one, and so on.
Then introducing variables (\ref{u_r}) for $n=0,1,2,3$ we arrive at system (\ref{chain_4}). 

\end{example}

\section{Conclusions}

Integrable differential-difference exponential type systems are presented corresponding to any simple or affine Lie algebra. In fact they are discrete analogues of the well-known Toda field equations. In the literature such kind systems are called lattice Toda field equations (see, for instance, \cite{KNS}, \cite{IH}). Our systems are different from those found earlier in \cite{IH} except the case $A_N$. The system corresponding the algebra $A_N$ called two-dimensional Toda molecule equation has been found in \cite{Hirota}.

We suggested a formal procedure which assigns to any Cartan matrix a system of differential-difference equations. Actually this correspondence has an important algebraic interpretation. Recently it was observed that any differential-difference system of the form 
$$
u^i_{1,x}=f^i(\mathbf u,\mathbf u_x,\mathbf u_1),\quad j=1,...,k, \mathbf u=(u^1,u^2,...,u^k), \mathbf u=\mathbf u(n,x), \mathbf u_1=\mathbf u(n+1,x)
$$ 
admits characteristic Lie algebras in $x$ and $n$ directions. It is shown in the article that the characteristic $x$-algebras of the systems (\ref{sysu}) are isomorphic to the Borel subalgebras of the simple Lie algebras related to the corresponding Cartan matrices.

\section*{Acknowledgments}
This work is partially supported by the Scientific and Technological Research Council of
Turkey (TUBITAK) grant 209 T 062, Russian Foundation for Basic Research (RFBR) grants $\#$ 10-01-91222-CT-a, $\#$ 11-01-97005-r-povoljie-a, and $\#$ 10-01-00088-a.

\end{document}